\documentclass[11pt]{article}

\usepackage[margin=1in]{geometry}
\usepackage[T1]{fontenc}
\usepackage{lmodern}
\usepackage{amsmath}
\usepackage{amssymb}
\usepackage{mathtools}
\usepackage{microtype}
\usepackage{enumitem}
\usepackage[round,authoryear]{natbib}
\usepackage{hyperref}

\hypersetup{
  colorlinks=true,
  linkcolor=blue,
  citecolor=blue,
  urlcolor=blue
}

\newcommand{\doi}[1]{\href{https://doi.org/#1}{doi:#1}}

\title{Does the Actuality of Life Favor Many Actual Histories?}
\author{Jonathan Baxter}
\date{May 2, 2026}

\begin{document}

\maketitle

\begin{abstract}
This essay develops a simple conditional argument for many-history
ontologies from the fact that life exists. The question is not why our
exact evolutionary sequence occurred, but why reality contains the
broad class of life-bearing histories at all. If such histories are
sufficiently rare, then a single-history ontology makes the existence
of life surprising, while an ontology containing many histories can
make it unsurprising. In the limiting case, life may be vanishingly
unlikely in any one history while almost certain to occur somewhere if
many histories are actualized. Everettian quantum mechanics is a
natural candidate for this kind of ontology because its multiplicity
is not introduced for anthropic purposes, but arises from taking
universal unitary quantum mechanics seriously. The Fermi paradox is
discussed as pressure against the claim that life is common within a
single history.
\end{abstract}

\section{Introduction}

Why does the actual world contain life? The question is not why human
beings exist, or why this exact evolutionary sequence occurred, but
why reality contains any life-bearing history at all: any physical
history in which matter becomes organized into self-maintaining,
reproducing, evolving systems.

One possible answer is that life is not surprising. Perhaps the
fundamental laws of physics strongly favor stable matter, chemistry,
planets, replication, and evolution. Perhaps, once a universe has the
right physical structure, life-bearing histories are common. If that
is true, then the existence of life gives us little reason to prefer
an ontology with many actual histories over an ontology with only one.

But another possibility is that life-bearing histories are extremely
rare. The rarity might enter at many levels. Maybe the fundamental
laws and constants are only narrowly compatible with stable
complexity. Maybe suitable planets are rare. Maybe abiogenesis is
fantastically unlikely even on suitable planets. Maybe simple life is
common but open-ended biological complexity is rare. The bottleneck
could occur at the lower level of fundamental physics, or higher up in
the contingent details of cosmology, chemistry, abiogenesis, and
evolution.

The empirical status of these questions remains unsettled. We have
only one known example of abiogenesis, observed from within a
life-bearing history, so the rarity of life-bearing histories cannot
be inferred directly from Earth's history. The early emergence of life
on Earth may suggest that abiogenesis is not extremely difficult, but
that inference is complicated by selection effects and prior
assumptions~\citep{SpiegelTurner2012,Kipping2020}.

This essay develops a conditional argument: if life-bearing histories
are extremely rare in a single-history ontology, then the actuality of
life supports an ontology in which many physical histories are
actualized.

Everettian quantum mechanics, often called many-worlds, is a
particularly interesting candidate for such an ontology. Its
multiplicity is not introduced merely to solve an anthropic puzzle. It
arises from taking the universal quantum state seriously and refusing
to add a physical collapse: a special process by which one quantum
outcome becomes actual while the others disappear~\citep{Everett1957}.

On the Everettian view, the alternatives represented in the quantum
state are not merely possible ways things might have gone. Under the
right conditions, they correspond to physically real branches of the
world~\citep{Wallace2013}. This makes Everett relevant to the question
of life because it changes what it means for a rare physical history
to be actual. In a single-history ontology, if life-bearing histories
are extremely rare, then our one actual history was very unlikely to
be life-bearing. In an Everettian ontology, by contrast, a rare
life-bearing history need not be the one selected history. It may be
one among many physically real branches.\footnote{The argument is
related to Wilson's treatment of multiverse hypotheses as undercutting
defeaters for fine-tuning arguments~\citep{Wilson2020}. Wilson's focus
is on whether Everettian or other multiverse hypotheses undercut the
inference from life-permitting constants to design. The present
argument differs in emphasis: it is not primarily an argument about
design or fine-tuning, but a positive abductive argument from the
actuality of life-bearing histories to many-history ontology.}

The claim is conditional. If life-bearing histories are common, then
no appeal to many actual histories is needed. But if life-bearing
histories are rare, then the fact that reality contains one supports
an ontology that actualizes many histories rather than only one.

The argument has four parts. First, I distinguish a single-history
ontology from a many-history ontology. Second, I clarify the relevant
datum: not this exact evolutionary history, but the actuality of a
life-bearing history disclosed from within experience. Third, I
consider where the rarity of life might enter: in the laws, in
cosmological history, in abiogenesis, or in later biological
complexity. Finally, I address the main objections: anthropic
conditioning, Born measure, rival many-trial hypotheses, and the
possibility that life is not rare after all.

The conclusion is deliberately modest. If life-bearing histories are
rare under a single-history ontology, and if the actuality of a
life-bearing evidential standpoint is legitimate evidence rather than
something to be conditioned away, then many-history ontologies
receive abductive support. Everett is the most interesting such
ontology because its multiplicity is already suggested by quantum
mechanics itself.

Before developing that argument, we need to clarify what ``one actual
history'' and ``many actual histories'' mean.

\section{One actual history versus many actual histories}

By a history, I mean a complete physical story of how the world
unfolds: which stars form, which planets form, which chemical
reactions occur, whether life begins, how evolution proceeds, and what
macroscopic events actually happen. A history is not just a momentary
state. It is an entire trajectory of events.

The argument depends on a contrast between two ways reality might be
structured. On one view, reality contains only one actual
history. Many things could have happened, but only one sequence of
events really does happen. Other possibilities may be physically
possible, mathematically represented, or assigned probabilities, but
they are not actual. They are unrealized alternatives. On another
view, reality contains many actual histories. Different possible
sequences of events are not merely unrealized alternatives. They are
physically real in some sense.

This contrast becomes especially important in quantum
mechanics. Quantum mechanics characteristically represents a system as
involving several possible outcomes at once. For example, a
radioactive atom may be described as having one component in which it
decays and another component in which it does not decay. When that
atom interacts with a measuring device, the measuring device may
become correlated with these alternatives: one component of the total
state records decay, while another records no decay.

The question is: what does that quantum description mean physically?

A single-history view says that, when quantum theory presents several
possible outcomes, only one of them becomes part of actual
reality. The other outcomes may appear in the mathematics. They may
help determine probabilities. They may describe things that could have
happened. But they do not actually occur. So, if a radioactive atom
either decays or does not decay, a single-history view says that one
of those things really happens and the other does not. If the decay
later affects a mutation, an organism, or an evolutionary path, then
only one resulting biological history is actual. Reality follows one
path through the space of physical possibilities.

Examples of single-history approaches include textbook collapse or
Copenhagen-style formulations, objective-collapse theories such as
GRW~\citep{GhirardiRiminiWeber1986}, hidden-variable theories such as
de Broglie--Bohm pilot-wave theory~\citep{Bohm1952a,Bohm1952b}, and
more recent stochastic approaches such as Barandes's
indivisible-stochastic formulation~\citep{Barandes2025}. These views
differ sharply in ontology and dynamics: GRW modifies quantum dynamics
with spontaneous collapses, Bohmian mechanics adds definite particle
configurations guided by the wavefunction, and Barandes recasts
quantum systems as indivisible stochastic processes in configuration
space. For present purposes, however, they share the relevant feature:
a single actual history is selected or realized, rather than all the
alternatives represented in the quantum state being equally actual.

A many-history view says something different. It says that the
different outcomes represented in the quantum state are not merely
possible outcomes, one of which is selected. Rather, when the
alternatives become separated into effectively non-interacting
branches, more than one outcome is physically real. In one branch, the
atom decays. In another branch, it does not. If those alternatives
later lead to different macroscopic events, then reality contains
multiple macroscopic histories.

The best-known many-history view is Everettian quantum mechanics. On
this view, the universal quantum state never collapses down to a
single outcome. It evolves according to the unitary dynamics of
quantum mechanics. What we call a measurement is not a special
physical process that selects one outcome and deletes the others. It
is a process by which different outcomes become correlated with
different branches of the world~\citep{Everett1957,Wallace2013}.

This does not mean that anything imaginable happens. The branches are
not arbitrary fantasies. They are constrained by the actual quantum
state, the actual laws of physics, and the dynamics of quantum
evolution. Nor does the ordinary Everettian picture require that
different branches have different fundamental laws or constants. At
the level of familiar quantum measurements and decohering macroscopic
events, the branches are different histories unfolding under the same
underlying physical laws.

In short:

\[
\text{Single-history ontology: many possible histories, but only one
  actual history.}
\]

\[
\text{Many-history ontology: many possible histories, and many actual
  histories.}
\]

This distinction matters for the argument from life. If life-bearing
histories are common, then one actual history may be enough. A
single-history universe would not need extraordinary luck to contain
life. But if life-bearing histories are extremely rare, then the
difference between one actual history and many actual histories
becomes important. A single-history ontology actualizes only one path
through physical possibility. If almost all such paths are lifeless,
then it is surprising that this one actual path contains life. A
many-history ontology changes the comparison. It does not need the
life-bearing history to be the one selected path. It needs only for
life-bearing histories to occur somewhere among the many actualized
histories.

The argument is therefore not that our exact world is
surprising. Every exact world-history would be highly specific. The
relevant question is whether the whole class of life-bearing histories
is rare. If it is, then a theory that actualizes many histories may
have an explanatory advantage over a theory that actualizes only
one. In rough terms, a single-history ontology faces the question,
``Why did the one actual history land in the rare life-bearing
class?'' A many-history ontology can answer, ``It did not have to be
the one selected history; life-bearing histories are actual because
many histories are actual.'' The rest of the essay develops that
thought more carefully, including the qualifications about measure,
anthropic conditioning, and the empirical rarity of life.

\subsection{A caveat about branching}

The phrase ``many histories'' should not be used too casually. In
Everettian quantum mechanics, there is a technical question about what
exactly counts as a branch.

A quantum state can be mathematically decomposed in many different
ways. So if one says that all branches are real, one must ask:
branches in which decomposition? Why treat the quasi-classical
branches, the ones containing stars, planets, organisms, records,
measurements, and ordinary macroscopic events, as the physically
relevant histories, rather than some mathematically different
decomposition of the same quantum state?

This is often called the preferred-basis problem. It is a real issue
for Everett. The usual Everettian answer appeals to
decoherence. Physical systems constantly interact with their
environments. Those interactions suppress interference between certain
components of the quantum state. The components that remain
dynamically stable are the ones that look approximately classical:
localized objects, stable records, definite-looking outcomes, and
macroscopic histories~\citep{Zurek2003}. \citet{Wallace2013}
summarizes the modern Everettian view as relying heavily on
decoherence to identify the quasi-classical structures that function
as branches, though this remains philosophically contested.

On this view, branches are not fundamental extra entities added to
quantum mechanics. They are emergent, approximate structures within
the universal wavefunction. A branch is not like a separate universe
floating somewhere else. It is a dynamically stable pattern in the
total quantum state.

For present purposes, the preferred-basis problem can be bracketed
without being ignored. \citet{HemmoShenker2022}, for example, argue
against treating decoherence alone as a complete solution. The
argument here assumes only the weaker claim that Everettian quantum
mechanics, if viable at all, supplies a robust multiplicity of
quasi-classical histories. That is enough for the present purpose. The
issue is not exactly how many branches there are, or exactly where one
branch splits from another. The issue is whether reality contains many
actual macroscopic histories rather than only one.

The argument is therefore conditional on that minimal Everettian
claim. If the usual decoherence-based picture is broadly correct, then
Everett supplies the kind of history-level multiplicity relevant to
the argument from life.

\section{Where could the rarity enter?}

The actuality of life might be surprising at several different levels.

At the lowest, most fundamental level, the laws and constants might be
finely tuned for complexity. If the strengths of forces, particle
masses, or cosmological parameters had been different, perhaps stable
matter, stars, heavy elements, or chemistry would not have
existed. The general thought that the basic structure of the physical
world depends delicately on a small number of microphysical constants
has a long history in anthropic reasoning~\citep{CarrRees1979}. If the
main improbability lies here, then Everettian branching within a fixed
low-energy physics may not be enough.

At a higher, more contingent level, perhaps the laws allow complexity,
but cosmological history must go a certain way. Stars must form. Heavy
elements must be produced. Planets must arise in suitable
environments. Usable sources of free energy must exist. If those
conditions are rare across possible histories under fixed laws, then a
many-history ontology becomes more relevant.

Higher up still, perhaps planets and chemistry are common, but
abiogenesis is extremely unlikely. The first self-maintaining,
reproducing systems may require an extraordinary sequence of chemical
accidents. If so, then even a universe with many planets might usually
remain lifeless. Existing evidence from Earth's early life does not
settle this issue. \citet{SpiegelTurner2012} argue that early
emergence is compatible with low abiogenesis rates once
observation-selection effects are included, while \citet{Kipping2020}
finds only modest support for rapid abiogenesis and some support for
rare intelligence.

If abiogenesis depends on such a sequence, then scientific explanation
may eventually bottom out in historical contingency. We might
reconstruct the chemical pathway by which life arose, while still
having to say that the required accidents happened. That would not
make the explanation unscientific, but it would leave the actuality of
a life-bearing history as a further explanatory question. Indeed, the
more the origin of life appears to depend on an extremely unlikely
sequence of contingent events, the more the actuality of a
life-bearing history itself becomes abductive evidence for an ontology
that actualizes many histories rather than only one.

A different bottleneck may lie not at the origin of life, but in the
transition from simple life to complex life. Maybe prokaryotic life is
common, while eukaryotes, multicellularity, nervous systems, or
open-ended ecological complexity are rare. That would make the
argument more specific, but the same structure would remain.

The key point is this: the actuality of life supports a many-history
theory only to the extent that life is rare at a level where that
theory supplies multiplicity.

If the bottleneck is history-level contingency under fixed laws,
Everett is directly relevant. If the bottleneck lies in the laws or
constants themselves, then ordinary Everettian branching under a fixed
low-energy Hamiltonian may not be enough by itself: it supplies many
histories governed by the same effective laws, not automatically many
sets of laws or constants. But that limitation should not be
overstated. In a deeper quantum-cosmological theory, the universal
quantum state might include superpositions over different vacuum
states, compactifications, field configurations, or effective
constants.

\citet{Susskind2003} presents the string-theory landscape as one
important setting in which a vast diversity of possible low-energy
vacua is taken seriously and explicitly connected to anthropic
reasoning. \citet{Shaya2025} proposes enlarging quantum-cosmological
configuration space to include theory-defining parameters themselves,
so that the universal wavefunction has support over worlds with
different effective physical laws.

The argument does not require these proposals to be correct. They
serve a narrower role: they illustrate how quantum superposition might
apply not just to events within a fixed universe, but also to the
structures that determine what kind of universe a branch contains. The
more careful claim is that Everett most directly helps with
history-level rarity under fixed physics, while Everett combined with
a landscape or quantum-cosmological framework may also bear on
law-level rarity.

\section{The relevant datum}

The argument should not begin from an imagined view from nowhere, as
though I first survey a space of possible worlds and then ask what
fraction, or what measure, of them contain life. That is not our
actual epistemic situation. I do not first inspect possible histories
and then infer that one of them contains organisms. I begin from
within this life-bearing history.

More carefully, the only thing I have absolutely direct evidence of is
experience itself. Experience is immediate. Everything else is
inferred. But some inferences are so basic and so overwhelmingly
supported that denying them would be less reasonable than accepting
them. From within experience, I have overwhelming evidence of a living
body, a biological environment, other organisms, reproduction,
evolution, and a world structured by life. Life is not given with the
same indubitability as experience, but it is part of the world
disclosed through experience with overwhelming force.

That is the epistemic starting point. The next question is how to
describe the datum for purposes of explanation. It would be too weak
to say merely that some possible world contains life. It would also be
too strong to require an explanation of this exact biological history
in every detail. The relevant datum lies between these extremes:

\[
E = \text{a life-bearing evidential situation is actual.}
\]

That formulation matters. The existence of life is not an optional
theoretical posit added after the fact. It is part of the evidential
standpoint from which theory choice begins. I am not starting from
nowhere and asking where I should expect to land. I am starting from
the brute fact that this life-bearing evidential situation is
actual. The question is which ontology makes that actuality less
surprising.

This also explains why consciousness matters without making
consciousness the whole point. Consciousness is epistemically
privileged because experience is where evidence begins. But the thing
to be explained is broader than consciousness: the actuality of a
physical history containing life, including cells, metabolism,
reproduction, ecosystems, organisms, and eventually beings capable of
reflecting on the fact that such a history is actual.

Once the datum is framed this way, the relevant contrast is not
between this exact history and all others. Every exact world-history
would be highly specific. The relevant contrast is between histories
that contain life and histories that do not. The question is whether
the broad class of life-bearing histories is rare.

That distinction is crucial. The actuality of a life-bearing history
could be surprising in the way a royal flush is surprising. The
actuality of this exact biological history would be merely surprising
in the way every exact shuffle is surprising.

\section{The conditional argument}

The core argument can now be stated more precisely.

Suppose that, under the actual laws of physics and the theory's
natural measure over possible histories, life-bearing histories have
extremely small measure. Then a single-history ontology faces a
straightforward problem: it actualizes only one history. If almost all
of the measure is concentrated on lifeless histories, then it is
surprising that the one actual history contains life.

A many-history ontology changes the picture. If many physically
possible histories are actualized, then life-bearing histories need
not be the one selected history. They need only occur somewhere in the
physically real structure of histories.

In Everettian quantum mechanics, this becomes especially interesting.
If the universal quantum state contains branches corresponding to
life-bearing histories, then those histories are not merely possible.
They are actual branches of the total physical state. A
single-history theory may assign such histories a very low probability
of becoming actual. An Everettian ontology says instead that they are
actual, provided they occur in the branching structure at all.

The conditional argument can therefore be stated as follows:

\begin{enumerate}[label=\arabic*.]
\item We find ourselves in a life-bearing history.

\item If life-bearing histories have extremely small measure in a
  single-history rival's probability distribution over histories, then
  single-history ontologies make the actuality of life surprising.

\item Everettian quantum mechanics, if interpreted realistically and
  without collapse, actualizes many quasi-classical histories rather
  than one.

\item Therefore, if life-bearing histories are sufficiently rare, the
  actuality of life provides evidence for Everettian or otherwise
  many-history ontologies.
\end{enumerate}

The conclusion is conditional and abductive. Its force depends on
whether life is actually rare.

\section{Objections and qualifications}

The argument just stated is conditional in several ways. Three
objections are especially important. First, anthropic reasoning
challenges whether the actuality of life should be treated as evidence
rather than conditioned on from the beginning. Second, Everettian Born
measure raises the worry that the mere existence of low-weight
life-bearing branches cannot do the explanatory work required. Third,
even if the argument succeeds, it does not uniquely favor Everett,
since other many-trial hypotheses may supply similar multiplicity.
We address those three issues in turn.

\subsection{The anthropic objection}

A standard reply is that we could not observe a lifeless world. Of
course we find ourselves in a life-bearing history, because only
life-bearing histories contain observers who can ask the question.

That response is partly right. There is a selection effect. We should
not be surprised that our evidence comes from a world compatible with
the existence of evidence-gatherers. Observation-selection effects are
central to anthropic reasoning~\citep{Bostrom2002}. They also connect
to the self-indication assumption, according to which one's existence
can favor hypotheses containing more observers, though that principle
remains controversial~\citep{BostromCirkovic2003}.

But the selection effect does not automatically dissolve the question.
It explains why we should not expect to observe a lifeless world. It
does not, by itself, explain why any life-bearing evidential situation
is actual in the first place.

The dispute is about whether life-bearing actuality should be treated
as evidence or simply conditioned on from the beginning. One position
says that, once we condition on the existence of observers, the
presence of life is no longer evidentially relevant. The opposing
position says that this conditions away the very thing at issue. The
existence of a life-bearing standpoint is itself part of the data. A
theory that makes such standpoints fantastically unlikely should be
penalized; a theory that makes them less surprising should be favored.

The argument in this essay assumes the second approach. It treats the
actuality of a life-bearing history as evidentially relevant. The rest
of the argument therefore depends on rejecting a fully deflationary
anthropic stance.

\subsection{Born measure and the ``too cheap'' objection}

The most serious Everett-specific objection concerns Born measure. In
Everettian quantum mechanics, branches are not normally counted
equally. Branch weights play the role that probabilities play in
ordinary quantum predictions. But because all outcomes with nonzero
amplitude occur, the ordinary meaning of probability is not
straightforward. This is the probability problem: how can the Born
rule make sense in a deterministic theory in which all possible
outcomes occur? Wallace treats this as one of the central challenges
for Everettian quantum mechanics~\citep{Wallace2013}.

The objection is powerful because Everettian confirmation cannot
simply ignore measure. A low-weight branch containing some observation
should not automatically count as strong support merely because that
branch exists. Otherwise Everett would make every low-weight anomaly
evidentially cheap. If life-bearing branches have extremely small Born
measure, then one might object that Everett has not really made life
less surprising. It has only said that some tiny-weight branch
contains life.

This objection parallels the anthropic objection, but at the level of
branch weight rather than observation selection. The tempting reply is
to say that, if life-bearing branches have tiny Born measure, then I
should not expect to find myself in one. That may be right if the
question is one of self-location among branches. \citet{SebensCarroll2018},
for example, argue that self-locating credence in Everettian quantum
mechanics should follow the Born rule rather than treating branches as
equiprobable.

But the present argument is not primarily a self-location argument. It
is not asking where a randomly located observer should expect to find
herself. It is asking whether the actuality of any life-bearing
evidential standpoint is less surprising under an ontology that
actualizes many histories than under one that actualizes only one. In
a single-history ontology, if life-bearing histories have tiny measure,
then probably no life-bearing history is actual. In an Everettian
ontology, if life-bearing histories occur in the branching structure,
then they are actual, even if their total Born measure is small.

This does not make Born measure irrelevant. Born measure remains
central to ordinary prediction, typicality, and empirical confirmation
within Everett. It also blocks the crude claim that every nonzero
branch is equally explanatory. The narrower claim is only that Born
measure does not obviously exhaust the evidential question at stake
here. If the question is, ``What should a typical observer expect to
see?'' then Born measure is unavoidable. If the question is, ``Why is
there a life-bearing evidential situation at all?'' then Born measure
limits the argument but does not by itself defeat it.

\subsection{Other many-trial hypotheses}

The conclusion, if sound, is not uniquely Everettian.

A cosmological multiverse might also supply many opportunities for
life. So might an infinite spatial universe, a cyclic cosmology, or
some deeper physical theory with many domains governed by different
effective parameters. If any of these theories actualize many
life-attempting histories, they could receive similar
support. Tegmark's multiverse hierarchy is useful here because it
distinguishes different kinds of multiplicity: distant spatial
regions, inflationary regions with different effective constants,
Everettian quantum branches, and more radical mathematical
variation~\citep{Tegmark2009}.

The conclusion is therefore broader than Everett: rare life would
favor many actualized trials over one actualized trial.

Everett is special because it is not an ad hoc multiplier. Its
multiplicity arises from a serious interpretation of an already
successful physical theory. It does not say, ``Let us invent many
worlds so life is less surprising.'' It says, ``If quantum mechanics
applies universally and there is no collapse, then the physical state
already contains many quasi-classical branches.''

That gives Everett a privileged role among many-history views without
making it the exclusive candidate.

\section{The single-history challenge and the Fermi paradox}

A defender of a single-history ontology can challenge the argument at
its most important empirical premise: perhaps life-bearing histories
are not actually rare. There are two main ways to press this
challenge. First, perhaps physics itself favors life. Maybe stable
matter, chemistry, planetary systems, usable sources of free energy,
and eventually abiogenesis are natural outcomes of the laws. On this
view, life is not a cosmic accident. Once the right physical structure
exists, life-bearing histories may be common enough that a single
actual history is not surprising.

Second, even if life is unlikely at any particular site, perhaps a
single actual universe contains enough ordinary trials. The universe
is not one laboratory flask. It may contain vast numbers of planets,
moons, oceans, hydrothermal systems, impact environments, and chemical
experiments across billions of years. Even if abiogenesis is rare per
site, the number of sites may be large enough that life is likely
somewhere in one actual history.

Both replies face the same observational pressure: if life is common
enough to make our existence unsurprising in a single actual history,
why have we seen no clear evidence of it elsewhere? This is the
Fermi-paradox pressure. Classically, the paradox concerns the absence
of evidence for extraterrestrial technological civilizations, despite
the apparent likelihood that such civilizations might exist;
\citet{Hart1975} argued that the absence of extraterrestrials on Earth
requires explanation if technologically capable civilizations are
common, and \citet{Brin1983} later surveyed the ``Great Silence''
controversy. The point here is broader but weaker. The silence is not
direct evidence that abiogenesis is rare. It is strongest as evidence
against abundant detectable technological life, and only indirect
pressure against the broader claim that life itself is common. Still,
it prevents the single-history ``life is probably common'' reply from
being automatic.

This point can be connected directly to the many-history argument. Let
\(K\) be the number of independent origins of life within a single
quasi-classical history. Here \(K\) counts independent origins, not
merely multiple inhabited locations. If life arises once and then
spreads by panspermia or technological dispersal, that is still
\(K=1\) for present purposes. We can then ask how the measure is
distributed among histories with no life, exactly one independent
origin of life, and two or more independent origins:
\[
\mu(K=0), \mu(K=1), \mu(K\geq 2).
\]
The relevant question is not only whether life occurs somewhere in the
total structure of reality. It is also whether a given life-bearing
history should be expected to contain exactly one independent origin
of life or multiple independent origins.

Suppose a history contains \(N\) possible sites or opportunities for
life, and each site has a small independent probability, or
branch-measure contribution, \(p\), of producing life. Then \(K\) is
binomially distributed. When \(N\) is large and \(p\) is small, with
\(\lambda=Np\), this is well approximated by a Poisson distribution:
\[
\mu(K=k) \approx e^{-\lambda}\frac{\lambda^k}{k!}.
\]
Expanding to second order in \(\lambda\), we have:
\[
\mu(K=0)\approx 1-\lambda+\frac{\lambda^2}{2},
\]
\[
\mu(K=1)\approx\lambda-\lambda^2,
\]
and
\[
\mu(K\geq 2)\approx\frac{\lambda^2}{2}.
\]
Unconditionally, histories with two or more independent origins
are second-order in \(\lambda\), while histories with exactly one
origin are first-order.

Now condition on being in a life-bearing history at all:
\[
K \geq 1.
\]
Then:
\[
\mu(K=1 \mid K\geq 1)=\frac{\lambda e^{-\lambda}}{1-e^{-\lambda}},
\]
while:
\[
\mu(K\geq 2 \mid K\geq 1)=\frac{1-e^{-\lambda}(1+\lambda)}{1-e^{-\lambda}}.
\]
For \(\lambda \ll 1\), these become:
\[
\mu(K=1 \mid K\geq 1) \approx 1-\frac{\lambda}{2},
\]
and:
\[
\mu(K\geq 2 \mid K\geq 1) \approx \frac{\lambda}{2}.
\]
If life is sufficiently rare, then given that we are in a life-bearing
history, we should expect that history to contain exactly one
independent origin of life. This conditional result applies equally to
single-history and many-history ontologies. The relevant contrast is
not about solitude conditional on life, but about the actuality of
life in the first place. In a single-history ontology, rare life makes
the actuality of any life-bearing history surprising; in a
many-history ontology, life can be actual somewhere while most
life-bearing histories still contain only one independent origin.

The structure of the argument is therefore as follows. Single-history
plus rare life predicts apparent solitude conditional on life, but
makes the actuality of life surprising. Single-history plus common
life makes the Fermi-paradox pressure sharper. Many-history plus rare
life offers a third possibility: life is rare enough that our apparent
solitude is unsurprising, while still being actual somewhere because
many histories are actualized.

The evidential force of this argument is limited in two ways. First,
the Fermi paradox may tell us more about the rarity of technological
civilizations than about the rarity of microbial life; it also depends
on assumptions about detectability, timescales, communication,
colonization, extinction, and search coverage. The point is not to
infer \(\lambda \ll 1\) merely from the fact that we have observed no
second origin. Rather, the point is conditional: if independent
origins are sufficiently rare, then the absence of a second origin is
not surprising, and does not count against a many-history explanation
of why life is actual somewhere. Second, the Poisson model assumes
approximate independence. If there is some global feature of a history
that makes life easy everywhere once it is possible anywhere, then the
probability of multiple origins need not be suppressed in the simple
way described above.

But if the relevant bottleneck is local and independent, such as
abiogenesis at particular sites, then the scaling is exactly what one
should expect. One origin is rare. Two or more independent origins are
much rarer still.

Thus the Fermi paradox helps sharpen the conditional claim. The
argument is strongest if reality has the following structure:
\[
\mu(K=0)\approx 1,\quad
\mu(K=1)\ll 1,\quad
\mu(K\geq 2)\ll \mu(K=1).
\]

In words: almost all of the measure lies on lifeless histories; a tiny
fraction lies on histories containing life once; an even tinier
fraction lies on histories containing life multiple times. A
single-history ontology then makes our life-bearing situation
surprising. A many-history ontology can make life actual somewhere
while also making our apparent solitude unsurprising.

\section{Conclusion}

The question is not why this exact world exists. Every exact world
would be improbable. The question is whether life-bearing histories as
a class are rare. If they are not rare, then the actuality of life
provides little support for many-history ontology. A single-history
universe would be enough.

But if life-bearing histories are extremely rare, then the fact that
one is actual becomes evidentially significant. A single-history
ontology actualizes only one path through physical possibility. A
many-history ontology actualizes many. Everettian quantum mechanics is
a natural candidate for such an ontology because its multiplicity is
not invented for anthropic purposes; it emerges from taking universal
quantum mechanics without collapse seriously.

The Fermi paradox sharpens rather than weakens this point. If life is
common, our apparent solitude becomes more puzzling. If life is rare,
then a single-history ontology makes our existence more puzzling. A
many-history ontology offers a distinctive middle position: life can
be rare within almost every individual history, so that most
life-bearing histories contain only one independent origin, while
still being actual somewhere because many histories are actualized.

The resulting claim has clear limits. It is conditional on the
empirical rarity of life, on the viability of quasi-classical
branching, on a non-deflationary treatment of life-bearing actuality
as evidence, and on the claim that Born measure does not exhaust the
evidential question at stake. It also supports many-history ontology
more directly than Everett specifically. If a deeper
quantum-cosmological theory includes superpositions over effective
laws or parameters, the same style of reasoning may apply higher up
the explanatory chain, though that extension is more speculative than
ordinary Everettian branching.

The final claim is therefore narrow but substantive: if life-bearing
histories have sufficiently small measure in a single-history rival's
probability distribution over histories, then the actuality of life
gives abductive support to physically grounded many-history
ontologies.

\end{document}